\documentclass[aps,prl,twocolumn,superscriptaddress,balancelastpage,showpacs,reprint]{revtex4-1}

\pdfoutput=1

\usepackage{blindtext}
\usepackage{centernot}
\usepackage{graphicx}
\usepackage{amsmath,bbold}
\usepackage{times}
\usepackage{amssymb}
\usepackage{mathrsfs}
\usepackage{chemarr}
\usepackage{color}
\usepackage{url}
\usepackage{version}
\usepackage{tikz}
\usepackage[percent]{overpic}
\usepackage{bm}
\usepackage[export]{adjustbox}
\definecolor{linkcolor}{rgb}{0,0,0.6}

\newcommand{\ee}{\text{e}}

\newcommand{\bn}{\boldsymbol{\nabla}}
\newcommand{\rr}{\bm{\mathrm{r}}}
\newcommand{\uu}{\bm{\mathrm{u}}}
\newcommand{\mm}{\bm{\mathrm{m}}}
\newcommand{\J}{\bm{\mathrm{J}}}
\newcommand{\F}{\bm{\mathrm{F}}}

\usepackage{lipsum}

\usepackage[pdftex]{hyperref}

\begin{document}
  
%%%%%%%%%%%%%%%%%%%%%%%%%%

\title{Surface Tensions between Active Fluids and Solid Interfaces: bare vs dressed}

\author{R. Zakine}
\thanks{These authors contributed equally}
\affiliation{Universit\'e Paris Diderot, Laboratoire Mati\`ere et Syst\`emes Complexes (MSC), UMR 7057 CNRS, F-75205 Paris, France}

\author{Y. Zhao}
\thanks{These authors contributed equally}
\affiliation{School of Physics and Astronomy and Institute of Natural Sciences, Shanghai Jiao Tong University, Shanghai 200240, China}
\affiliation{Universit\'e Paris Diderot, Laboratoire Mati\`ere et Syst\`emes Complexes (MSC), UMR 7057 CNRS, F-75205 Paris, France}

\author{M. Kne\v zevi\'c}
\affiliation{Institut f\" ur Theoretische Physik, Technische Universit\" at Berlin, Hardenbergstra\ss e 36, D-10623 Berlin, Germany}

\author{A. Daerr}
\affiliation{Universit\'e Paris Diderot, Laboratoire Mati\`ere et Syst\`emes Complexes (MSC), UMR 7057 CNRS, F-75205 Paris, France}

\author{Y. Kafri}
\affiliation{Department of Physics, Technion, Haifa 32000, Israel}

\author{J. Tailleur}
\affiliation{Universit\'e Paris Diderot, Laboratoire Mati\`ere et Syst\`emes Complexes (MSC), UMR 7057 CNRS, F-75205 Paris, France}

\author{F. van Wijland}
\affiliation{Universit\'e Paris Diderot, Laboratoire Mati\`ere et Syst\`emes Complexes (MSC), UMR 7057 CNRS, F-75205 Paris, France}

\date{\today}

\begin{abstract}
We analyze the surface tension exerted at the interface between an active fluid and a solid boundary in terms of tangential forces. Focusing on active systems known to possess an equation of state for the pressure, we show that interfacial forces are of a more complex nature. Using a number of  macroscopic setups, we show that the surface tension is a combination of an equation-of-state abiding part and of setup-dependent contributions. The latter arise from generic setup-dependent steady currents which "dress" the measurement of the "bare" surface tension. The former shares interesting properties with its equilibrium counterpart, and can be used to generalize the Young-Laplace law to active systems. We finally show how a suitably designed probe can directly access this bare surface tensions, which can also be computed using a generalized Virial formula.
\end{abstract}

\maketitle 

Active particles are driven out of equilibrium through a dissipative
exchange of energy and momentum with their environment, which endows
them with anomalous thermomechanical
properties~\cite{di2010bacterial,sokolov2010swimming,mallory2014anomalous,takatori2014,yang2014,solon_pressure15,bialke15,sandford2017pressure,junot2017active,vizsnyiczai2017light,fily17,patch2018}. Among
these, the pressure exerted by active systems on their confining
vessels has recently attracted a lot of
interest~\cite{mallory2014anomalous,takatori2014,yang2014,solon_pressure15,solon2015pressure,nikola16,speck2016,marconi16,sandford2017pressure,junot2017active,fily17,ginot2018sedimentation,falasco2016}. In
particular, the existence of an equation of state for the pressure was
established for a large subclass of these non-equilibrium and
momentum-non-conserving systems. The defining feature which allows for an equation of state is a
self-propulsion force whose dynamics is independent of the positional
and angular degrees of freedom~\cite{fily17}. This encompasses standard
models of active particles, such as run-and-tumble particles,
active Brownian particles (ABPs), and active Ornstein-Uhlenbeck
particles (AOUPs), both non-interacting and in the presence of
pairwise conserving forces. For such systems, which are the focus of
this article, macroscopic mechanical properties echo the equilibrium
case with, for instance, equality of pressures in coexisting
phases~\cite{solon2015pressure}; This raises the hope of a generalized
thermodynamics~\cite{ginot2015nonequilibrium,solon2018generalized,rodenburg2017van,solon18}.

While pressure controls the bulk thermodynamic state of a system in
equilibrium, its interfacial properties rely on surface tension,
perhaps the most elusive thermomechanical
quantity whose microscopic origin has been
the topic of long-standing debates~\cite{berry71,marchand11,rowlinson82}. Surface tension
indeed controls a wealth of phenomena, from the demixing of binary
mixtures to the wetting of surfaces, and the instability of thin films
and
jets~\cite{de2013capillarity,bonn2009,eggers2008,craster2009}. Given
the atypical properties of active matter at interfaces, from the
accumulation at
boundaries~\cite{elgeti2009self,tailleur2009sedimentation,elgeti2013wall}
to the wetting of soft gels by swarming
bacteria~\cite{kearns2010,darnton2010,hennes2017active}, the study of
surface tension in active systems has naturally attracted
attention~\cite{bialke15,paliwal17}. In particular, the surface
tension of active particles at a liquid-gas interface in a system
undergoing motility-induced phase separation has been measured using
an expression derived by Kirkwood and Buff for Hamiltonian
systems~\cite{kirkwood49,buff55}. Despite stable
interfaces, this liquid-gas surface
tension was, somewhat surprisingly, found to be
negative~\cite{bialke15,solon18,patch2018}. Its thermodynamic role has nevertheless been
confirmed: it controls the pressure drop through the boundary of a
circular droplet of radius $R$ through a Laplace pressure given by
$\Delta P = \frac{\gamma}R$~\cite{solon18}. This combination of
surprising and familiar aspects of interfacial physics enjoins us to
clarify the microscopic origin of surface tension in these active
systems and in particular its mechanical implications, which have been
little discussed so far. Accounting for capillary and wetting
phenomena in active matter is a formidable program which we attack by
elucidating \textit{the interfacial properties of active fluids in contact
with a solid boundary}. For simplicity, we work in two dimensions
throughout the article, the generalization to higher dimensions being
straightforward.

\begin{figure}
  \begin{tikzpicture}
    \path (-4,0) node {\includegraphics[totalheight=2.75cm]{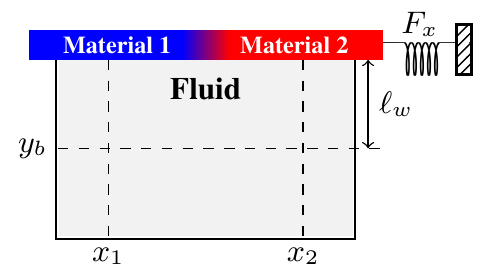}};
    \path (0.5,0) node {\includegraphics[totalheight=2.75cm]{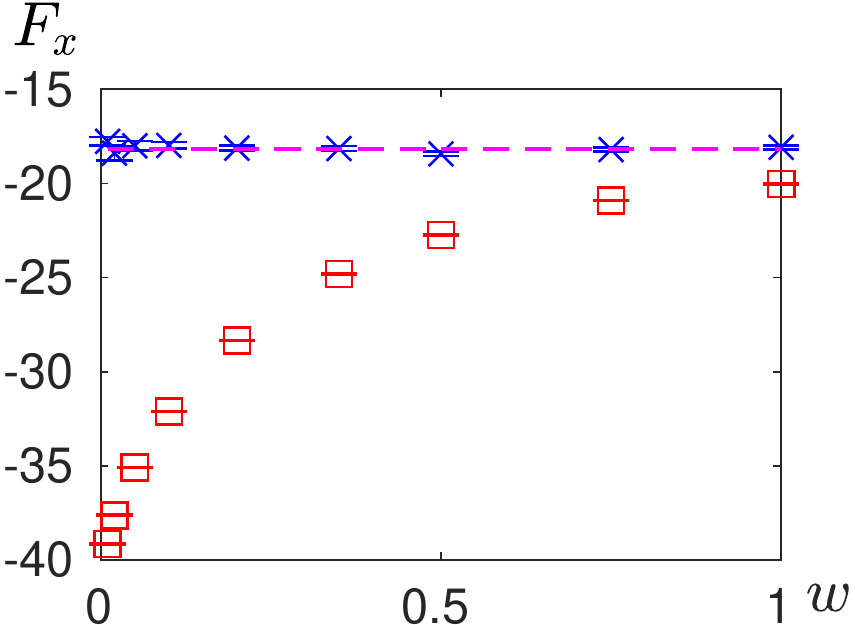}};
    \draw (-4.5,1) node[anchor=south west] {\bf a)};
    \draw (.25,1) node[anchor=south west] {\bf b)};
  \end{tikzpicture}
  \caption{{\bf a)} Modified Langmuir setup: a fluid is confined by a
    mobile upper boundary made of two different materials, modelled by
    a potential $V(x,y)=\lambda(x)(y-y_w)^4$ for $y>y_w$. The function
    $\lambda(x)=\lambda_1+\frac{\lambda_2-\lambda_1}{2}(1+\tanh\frac{x-l}{w})$
    interpolates smoothly between the two material stiffnesses
    $\lambda_1$ and $\lambda_2$. {\bf b)} Total tangential force $F_x$
    exerted on the upper wall as a function of the width $w$ of the
    junction. For an equilibrium system (blue symbols), $F_x$ is
    independent of $w$ and given by $\Delta \gamma$, Eq.~\eqref{eq:deltagammaEOS} (dashed line). In stark contrast, for
    the active system (red symbols), $F_x$ depends on $w$. Parameters:
    $\lambda_1=5$, $\lambda_2=0.05$, $l=L_x/2$. Equilibrium Brownian
    simulations: $D=8.2$, $L_x=2L_y=80$. ABPs:
    $v=5$, $D_r=1$, $L_x=2L_y=1280$, $\ell_w=10$. Throughout the article, the bulk density is 1.}\label{fig:Langmuir}
\end{figure}

In the absence of an established thermodynamic route, it is natural to
revert to mechanics. The Langmuir balance~\cite{langmuir1917} is a
standard tool to measure the tangential force exerted on an interface
between a fluid and a solid in equilibrium, and hence to access surface
tension: Take a mobile wall made of two different materials confining
a fluid in a cavity (see Fig~\ref{fig:Langmuir}-a); Moving the wall
changes the contact area between the fluid and each material; In
equilibrium, the corresponding tangential force exerted by the fluid
on the wall is thus given by the difference of surface
tensions between the fluid and
each material $\Delta \gamma= \gamma_2-\gamma_1$. This highlights an important property of equilibrium
systems: the tangential force is independent of the details of the
junction between the two walls, as illustrated in
Fig~\ref{fig:Langmuir}-b. Strikingly, similar computations carried out
for non-interacting ABPs depict a very different physics: the
tangential force now strongly depends on the details of the junction
between the two materials. The existence of an equation of state for
the pressure of an active fluid thus does not entail a similar
property for the fluid-solid surface tension.

In this article we indeed show that, in addition to an intrinsic,
\textit{bare}, equation-of-state-abiding contribution, the emergence
of junction-dependent currents leads to an extra contribution that
prevents the existence of a global equation of state for the
tangential force. The Langmuir setup of Fig.~\ref{fig:Langmuir} hence
only accesses a surface tension dressed by these junction-dependent
currents: the reading of the surface tension depends on the probe used
for its measurement. 
In addition to its intrinsic, junction-independent nature, the bare surface tension plays an important, versatile role in other surface phenomena, much like in equilibrium.  For example, we show how it controls the Young-Laplace corrections to the normal pressure felt by a circular confining cavity.
We thus suggest a second setting designed to make
the contribution of the currents vanish, hence providing a direct
mechanical measurement of the bare surface tension. Finally, a lot of
effort has been invested in the past to derive the pressure of active
fluids using a Virial
approach~\cite{yang2014,takatori2014,winkler2015virial,falasco2016,speck2016}.
Here we introduce a different Virial, which offers a macroscopic route
to the bare surface tension.

\textit{Current-induced tangential forces.}
To model the experiment depicted in Fig.~\ref{fig:Langmuir}-a, we follow
Navascu\'es and Berry~\cite{navascues77} and disregard the internal
dynamics of the wall, which we model as a confining potential smoothly
interpolating between the two materials. This allows us to focus on
the new features due to activity and leads to a straightforward
expression for the tangential force exerted by the particles on the
wall: 
\begin{equation}\label{eq:Fx}
  F_x=\int_{x_1}^{x_2} dx \int_{y_b}^\infty dy\,\rho(x,y) \partial_x V(x,y)
\end{equation}
where $\rho$ is the mean density of particles, $x_1$ and $x_2$ are abscissa
far away from the junction between the two materials, and $y_b$ is an
ordinate in the `bulk' of the system, at a
distance $\ell_w\equiv y_w-y_b$ from the upper confining wall (See Fig~\ref{fig:Langmuir}).  For an equilibrium ideal gas,
a direct computation leads to
\begin{equation}\label{eq:deltagammaEOS}
F_x=\gamma_2-\gamma_1\quad\mbox{with} \quad \gamma_k\equiv -\int_{y_w}^\infty dy\,kT \rho(x_k,y)\;.
\end{equation}
$F_x$ is indeed independent of the details of the junction between the
two walls. This is expected from thermodynamical considerations, from
which~\eqref{eq:deltagammaEOS} could have been
derived~\footnote{Consider the fluid in a rectangular cavity of linear
  sizes $L_x$ and $L_y$, bounded by a potential of stiffness
  $\lambda_i$. A direction computation leads to
  $\gamma=\left.\frac{\partial {\cal F}}{\partial
    L_x}\right|_{A=L_xL_y}=-\int_{y_w}^\infty kT \rho(x,y)$ where $x$
  is a position in the bulk of the system and ${\cal F}$ the free
  energy~\protect\cite{Nijmeijer1990}}.  

\begin{figure}
  \begin{tikzpicture}
    \path (0,0) node {\includegraphics[width=.49\columnwidth]{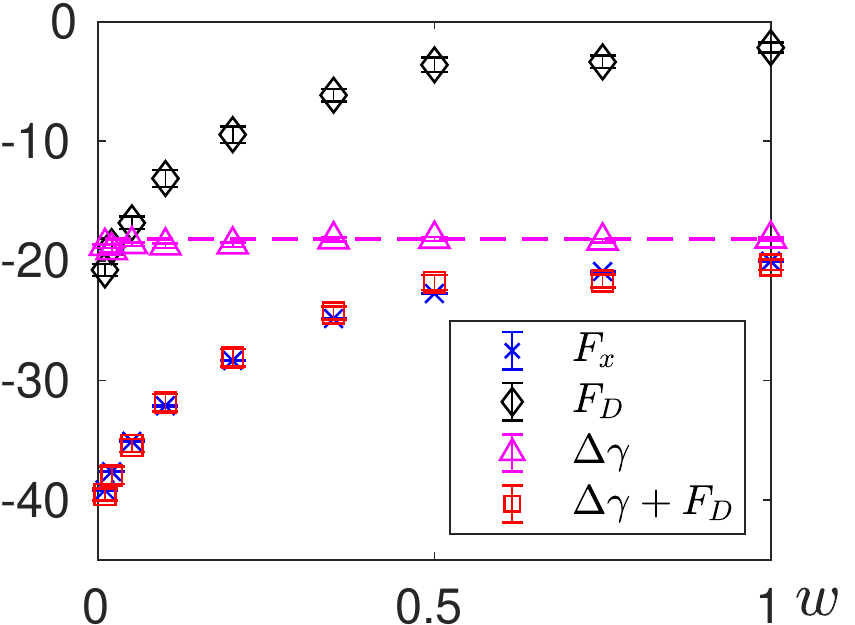}};
    \draw (-1.5,1.4) node[anchor=south west] {\bf a)};
  \end{tikzpicture}\hspace{-.35cm}
  \begin{tikzpicture}
    \path (0,0) node {\includegraphics[width=.49\columnwidth]{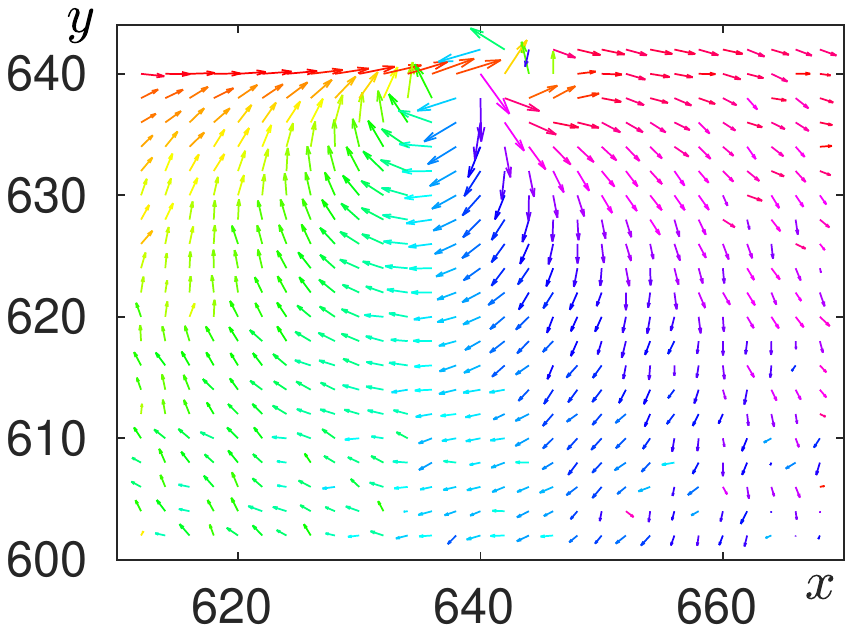}};
    \draw (-1.5,1.4) node[anchor=south west] {\bf b)};
  \end{tikzpicture}
  \caption{{\bf a)} Comparison between $F_x$ and the sum of
    the bare $\Delta \gamma$ and the junction-dependent drag force $F_D$. $\Delta \gamma$ coincides
    with its independent measurements using Eq.~\eqref{eq:baregamma}
  in homogeneous systems bounded by upper potentials
    $V_1=V(x\ll l)$ and $V_2=V(x\gg l)$ 
    (dashed line).  {\bf b)} Map of the current around the junction for $|\J|>10^{-6}$. The amplitude of the arrows are proportional to
    $\log|\J/10^{-6}|$; the color
    encodes its directions. In both figures, $\lambda_1=5$,
    $\lambda_2=0.05$, $l=L_x/2$, $v=5$, $D_r=1$, $L_x=2L_y=1280$. For the current field,
    $w=0.01$.}
\label{fig:currents_junction}
\end{figure}

To elucidate analytically the dependence of $F_x$ on the width of the
junction $w$ reported for ABPs in Fig.~\ref{fig:Langmuir}-b, we start
from the dynamics of $N$ non-interacting active particles:
\begin{align}\label{eq:dynamics}
  \dot \rr_i=v_0\uu_i-\mu\bn_{\rr_i}V(\rr_i),
\end{align}
where the active force on particle $i$ enters through $v_0
\uu_i /\mu$. We consider tumbles and rotational diffusion of the particle orientations $\uu_i=(\cos\theta_i,\sin\theta_i)$, and denote $\tau$ the persistence time. The generalization to
other models of active particles like AOUPs is straightforward. To
compute the tangential force on the upper wall, we introduce the
average density field $\rho(\rr)=\langle \sum_i
\delta(\rr-\rr_i)\rangle$, where the brackets represent averages over
the active-force statistics. It evolves according to the continuity
equation
\begin{align}\label{eq:dens}
  \partial_t \rho=-\bn\cdot \J,\qquad \J= v_0 {\bf m} - \rho \mu \bn V
\end{align}
where the local orientation field $\mm(\rr)=\langle \sum_i \uu_i
\delta(\rr-\rr_i)\rangle$ controls the contribution of self-propulsion
to the particle flow. In
steady-state (see, \textit{e.g.},
\cite{solon_pressure15})
\begin{align}
  m_\alpha=-\partial_\beta\left[ v_0 \tau \left(Q_{\alpha\beta}+\frac{\rho \delta_{\alpha\beta} }{2}\right)-{\mu}{\tau} m_\alpha\partial_\beta V\right]
  \label{eq:steady_state_magnetization}
\end{align}
where $Q$ measures the local nematic order through $Q_{\alpha
  \beta}(\rr)=\langle \sum_i (\uu_{i,\alpha} \uu_{i,\beta}
-\delta_{\alpha\beta} /2) \delta(\rr-\rr_i)\rangle$, with Greek
letters refering to space directions.
From Eqs.~\eqref{eq:Fx} and~\eqref{eq:dens}, the tangential force
$F_x$ exerted on the upper wall can be written as
\begin{equation}
F_x= F_D + \int_{x_1}^{x_2}\int_{y_b}^\infty \frac{v_0}{\mu} m_x(x,y)\, dxdy \label{eq:m+current}
\end{equation}
where $F_D$ is the total drag force experienced by particles around
the junction:
\begin{equation}\label{eq:FD}
  F_D=-\frac{1}{\mu}\int_{x_1}^{x_2}dx \int_{y_b}^\infty dy\, J_x=-\mu^{-1} \!\!\!\!\!\!\!\!\!\!\!\!\!\sum_{x_i\in[x_1,x_2];y_i>y_b} \!\!\!\!\!\!\!\!\!\!\!\!  \langle \dot \rr_i \cdot  \ee_x\rangle\;.
\end{equation}
Equation~\eqref{eq:m+current} has an appealing physical interpretation
in terms of force balance: the total active force exerted by the
particles, which is the second term on the
r.h.s. of~\eqref{eq:m+current}, is split between the force exerted on
the upper wall $F_x$ and the drag force $F_D$
exerted on the environment. Using
Eq.~\eqref{eq:steady_state_magnetization}, the isotropy of the bulk,
translational invariance far from the junction, and that
$\int_{y_b}^{\infty}dy Q_{xx}(x,y)=0$ for $x=x_1$ or
$x=x_2$~\cite{foot1}, the force balance equation~\eqref{eq:m+current}
can be rewritten as:
\begin{equation}
F_x= F_D + \gamma_2-\gamma_1  \label{eq:EOS+current}
\end{equation}
where we define $\gamma_k$ as
\begin{equation}\label{eq:baregamma}
\gamma_k=-\frac{v_0^2 \tau}{2\mu}\Big(\int_{y_b}^\infty dy\, \rho(x_k,y)-\int_{y_b}^{y_w}\, dy\, \rho_0\Big)\;.
\end{equation}
\begin{figure}
  \includegraphics[width=.42\columnwidth]{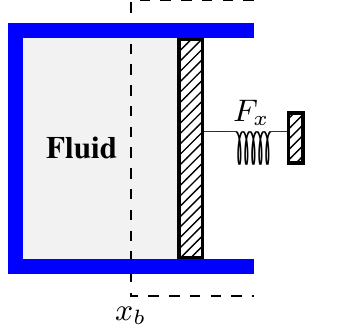}\hspace{.25cm}
  \raisebox{0.1cm}{\includegraphics[width=.48\columnwidth]{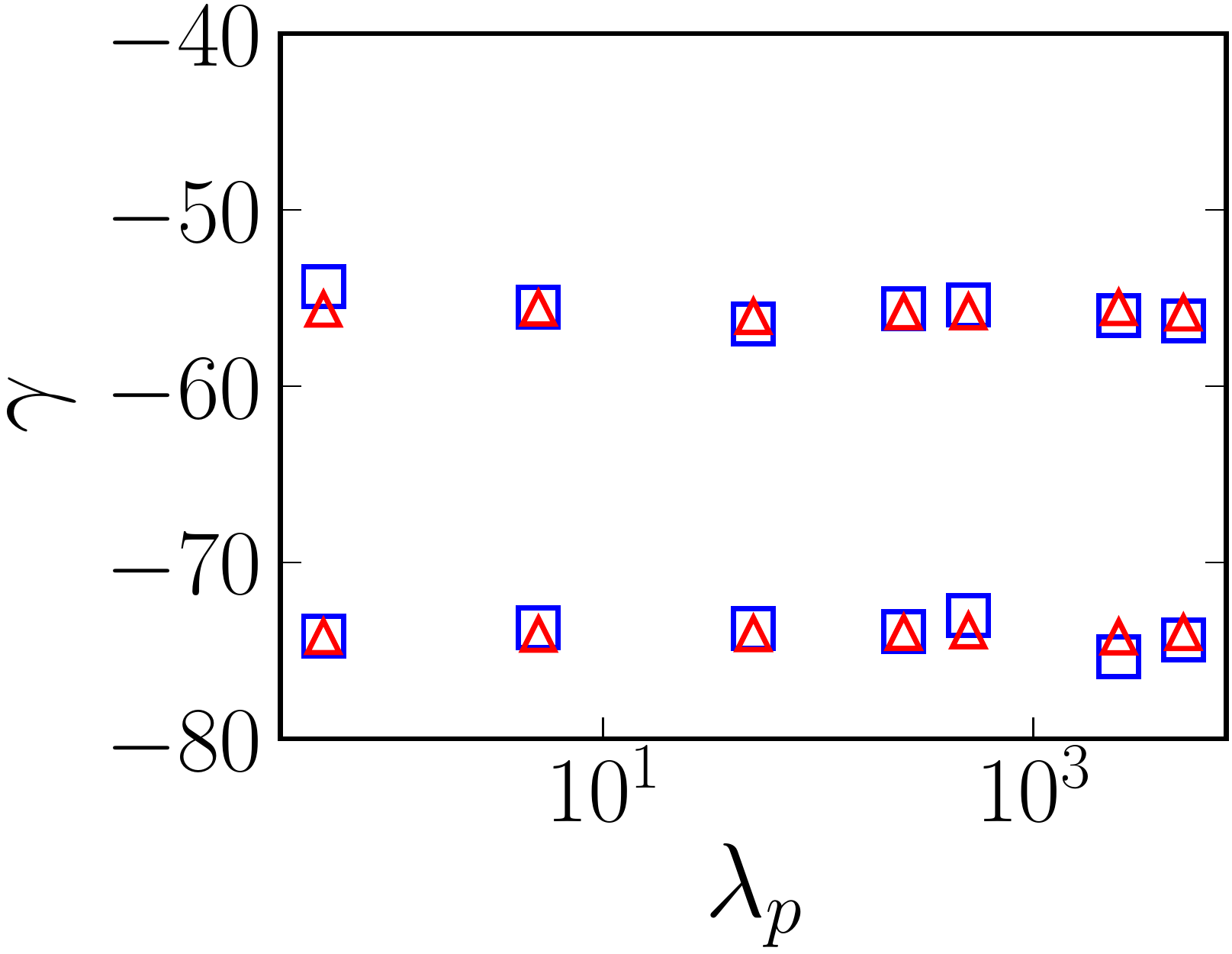}}
  \caption{{\bf Left:} Setup used to directly measure the bare surface tension. {\bf Right:} The measurement of $\gamma$ for two different confining materials are shown to be independent of the piston's stiffnesses $\lambda_p$. Measuring $\gamma$ from the force on the piston (blue squares) or from 
the `bulk correlator'~\eqref{eq:baregamma} evaluated at $x=x_b$ (red triangles) lead to consistent results. Parameters: $L_x=L_y=50$, $v_0=5$, $\tau=1$. Upper and lower results correspond to confining materials of stiffness $\lambda=5$ and $\lambda=0.05$, respectively. The distance between the two measurements corresponds to $\Delta \gamma$ in Fig.~\ref{fig:currents_junction}a.}
\label{fig:reduced_current_setup}
\end{figure}
Note that the final integral in Eq.~\eqref{eq:baregamma} cancels out in Eq.~\eqref{eq:EOS+current}; it ensures that $\gamma_k$ does not depend on $y_b$.
Equations~\eqref{eq:EOS+current} and~\eqref{eq:baregamma} form a
central result of this letter; They are exact in the macroscopic limit
$L_x,L_y, \ell_w\to \infty$ with $\rho_0=N/L^2$ finite and $\ell_p\ll
\ell_w \ll L_x,L_y$, where $\ell_p=v_0\tau$ is the persistence
length. The result has several implications. As in equilibrium,
$\Delta \gamma = \gamma_2-\gamma_1$ is a contribution to $F_x$ that is
independent of the details of the junction; Eq.~\eqref{eq:baregamma}
is the equation of state for $\gamma_k$. It is determined by the
excess density accumulated at the wall compared to the bulk
values. Note that $\gamma_k$ can be measured independently in a
translationally invariant system and thus solely depends on the
interaction between the active fluid and wall $k$. Conversely, $F_D$
strongly depends on the details of the junction: the presence of an
asymmetric potential combined with the nonequilibrium dynamics of the
active particles generates, as in a ratchet, steady currents which are
bound to be junction-dependent; the latter give rise to a non-zero
drag force which dresses $F_x$. Fig.~\ref{fig:currents_junction}-a
indeed shows that the dependence of $F_x$ on the width $w$ of the
junction is entirely due to the variations of $F_D$, whereas $\Delta
\gamma$ is independent of $w$, and can be obtained from independent
measurements of $\gamma_{1,2}$.  Figure~\ref{fig:currents_junction}-b
suggests that the junction generates a current field wich decays
slowly with the distance to the junction. This is consistent with
recent results on the power-law decay of currents generated by
asymmetric obstacles located in the bulk of active fluids~\cite{baek18}. Note that the
macroscopic limit defined above, in which $\ell_w \ll L_x,L_y$ ensures
that $\Delta \gamma$ is indeed independent of the junction. Otherwise,
if $\ell_w \sim L_x,L_y$, the measurement of $\Delta \gamma$ itself
could be modified by the currents. In our numerics, we found that
$\ell_w=2 \ell_p$ was sufficient for the system to be isotropic at
$y=y_b$ and gives a reliable measurement of the bare $\Delta\gamma$.

Finally, in the $\tau\to 0$ limit, with $v_0^2 \tau$ fixed, active
particles behave as equilibrium colloids at an effective temperature
$kT_{\rm eff}=v_0^2\tau/(2\mu)$ and Eq.~\eqref{eq:baregamma}
reduces to its equilibrium counterpart~\eqref{eq:deltagammaEOS}. The
partial cancellation of the two integrals in~\eqref{eq:baregamma} then
stems from the fact that, unlike active particles, passive ones have a
steady-state distribution which is a local function of the external
potential $V(x,y)$. In this limit, as for true equilibrium systems, no
currents survive in the steady state; $F_D$ vanishes and $F_x$ is then
a direct measurement of the bare, equation-of-state-abiding
contribution $\Delta \gamma$. 

Finally, the existence of an equation of state for surface tension extends to equilibrium fluids with interactions~\cite{navascues77}. Similarly, all our derivations extend to active fluids with pairwise forces at the cost of a lengthier expression for $\gamma$.

\textit{Bare surface tension and Laplace pressure corrections.}  The
use of surface tension is not limited, in equilibrium, to the
measurement of tangential forces. A famous example is that of the
Young-Laplace law which shows how surface tension impacts the pressure
on a circular interface. In our solid-liquid framework, this amounts
to computing the normal pressure exerted by an active fluid on a
circular confining wall of radius $R$, a problem which has been looked
at in the
past~\cite{mallory2014anomalous,yan2015force,sandford2017pressure}. The
pressure can be computed from $2\pi R P(R) = \int \rho(r) \partial_r V
r {\rm d}r {\rm d}\theta$. In the flux-free steady state observed in such a system, one has $\rho(r) \partial_r
V=v_0 m_r$. Since Eq.~\eqref{eq:steady_state_magnetization}
shows the local orientation to the divergence of a local
tensor~\cite{fily17}, the pressure is 
given,  in circular coordinates, by
\begin{align*}\label{eq:Laplacetemp}
  P(R)=&-\frac 1 R \int_0^\infty \partial_r [v_0\tau (Q_{rr}+\frac\rho
    2)-\mu\tau m_r\partial_r V]r\,dr\\&-\frac 1 R \int_0^\infty [v_0\tau
    (Q_{rr}-Q_{\theta\theta})-\mu \tau m_r \partial_r V]\,dr\;.
\end{align*}
Integrating by part the first term on the
r.h.s. leads to $P(R)=R^{-1} \int_0^\infty
v_0\tau (Q_{\theta\theta}+\rho/2)dr$. Adding and substracting the bulk pressure
$P_b=\rho v_0^2 \tau/(2\mu)$ then leads, in the large $R$ limit, to
\begin{equation}\label{eq:LYlaw}
  P(R)\simeq P_b -\frac 1 R \Big[\int_0^R \frac{v_0^2\tau \rho_0}{2\mu}\, dr - \int_0^\infty \frac{v_0^2 \tau \rho(r)}{2\mu}\, dr \Big]
\end{equation}
where $\rho_0$ is the density at $r=0$. Importantly,
Eq.~\eqref{eq:LYlaw} shows that the pressure can be written as a
generalized Young-Laplace law $P(R)= P_b-\frac{\gamma}{R}+o\big(\frac
1 R\big)$, with $\gamma$ identified as the bare surface tension
defined in~\eqref{eq:baregamma}, a connection not made
previously. Despite its dressing by currents in the microscopic
setting of Fig~\ref{fig:Langmuir}, the bare surface tension thus plays
an important role in active fluids. We now tackle the question of its
direct mechanical measurement.

\textit{Direct measurement of the bare surface tension.}
The currents arising in the Langmuir setup, which prevents the direct
measurement of the bare surface tension, are due to the
junction. Furthermore, the Langmuir setup only gives access to the
difference between the two fluid-solid surface tensions
$\Delta\gamma$. We now consider instead the setup depicted in
Fig.~\ref{fig:reduced_current_setup}, in which, as we show below, the
force exerted by the active fluid on a piston allows one to directly access the (bare) surface
tension between the fluid and the walls of the container. Using Eq~\eqref{eq:dynamics}, the total force $F_x=\int_{x_b}^\infty\int_{-\infty}^\infty dxdy
\rho(x,y) \partial_x V(x,y)$  exerted on the piston can be rewritten as
\begin{equation}\label{eq:Fx2eme}
  F_x=- \frac 1 \mu \langle {\sum_{i|x_i>x_b}} \dot x_i\rangle +  \int_{x_b}^\infty\int_{-\infty}^\infty dxdy\, \frac{v_0}{\mu}  m_x 
\end{equation}
Equation~\eqref{eq:steady_state_magnetization} then allows us to
simplify~\eqref{eq:Fx2eme} into
\begin{equation}
  F_x=F_D+\int_{-\infty}^\infty {\rm d}y\, \rho(x_b,y) \frac {v_0^2 \tau}{2 \mu}
\end{equation}
where $F_D$ is defined as the first term in the r.h.s. of~\eqref{eq:Fx2eme} and we have used that all fields vanish at infinity and that
$\int_{-\infty}^\infty Q_{xx}(x_b,y)dy=0$~\cite{foot1}. Introducing the bulk pressure
$P_b=\rho_0 v_0^2\tau/2\mu$, $F_x$ is finally given by
\begin{equation}\label{eq:gammapiston}
  F_x = F_D+ P_b L_y - 2 \gamma
\end{equation}
with $\gamma$ defined as in Eq~\eqref{eq:baregamma}, assuming for simplicity that the upper and lower
walls are identical. At first glance,
Eq.~\eqref{eq:gammapiston} seems to imply that, once again, the bare
surface tension is dressed by currents through the drag force
$F_D$. Note, however, that the average density of particles in the
half-plane defined by $x>x^*$ is constant, for all $x^*$, in
steady-state. Since $\dot \rho=-\nabla \cdot \J$, this implies that
$\int_{-\infty}^{\infty} J_x(x^*,y) {\rm d}y=0$. Integrating this from
$x^*=x_{b}$ up to infinity shows that $F_D=0$ in this setup: $F_x$ thus
provides a direct measurement of the bare surface tension between the
 fluid and the upper and lower walls, as shown in Fig~\ref{fig:reduced_current_setup}-right.

\textit{A Virial route to surface tension.} In equilibrium, the
mechanical route to surface tension can be circumvented by free energy
considerations, a path which is not accessible to active
systems. Another historical route to equations of states was proposed
by Clausius through the Virial. This has recently been applied successfully to determine the pressure of active fluids~\cite{takatori2014,yang2014,winkler2015virial,falasco2016,speck2016}
and we now extend this approach to surface tension.

\begin{figure}
  \includegraphics[width=0.45\columnwidth]{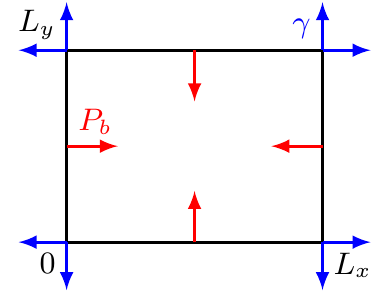}
\raisebox{-.3cm}{\includegraphics[width=0.45\columnwidth]{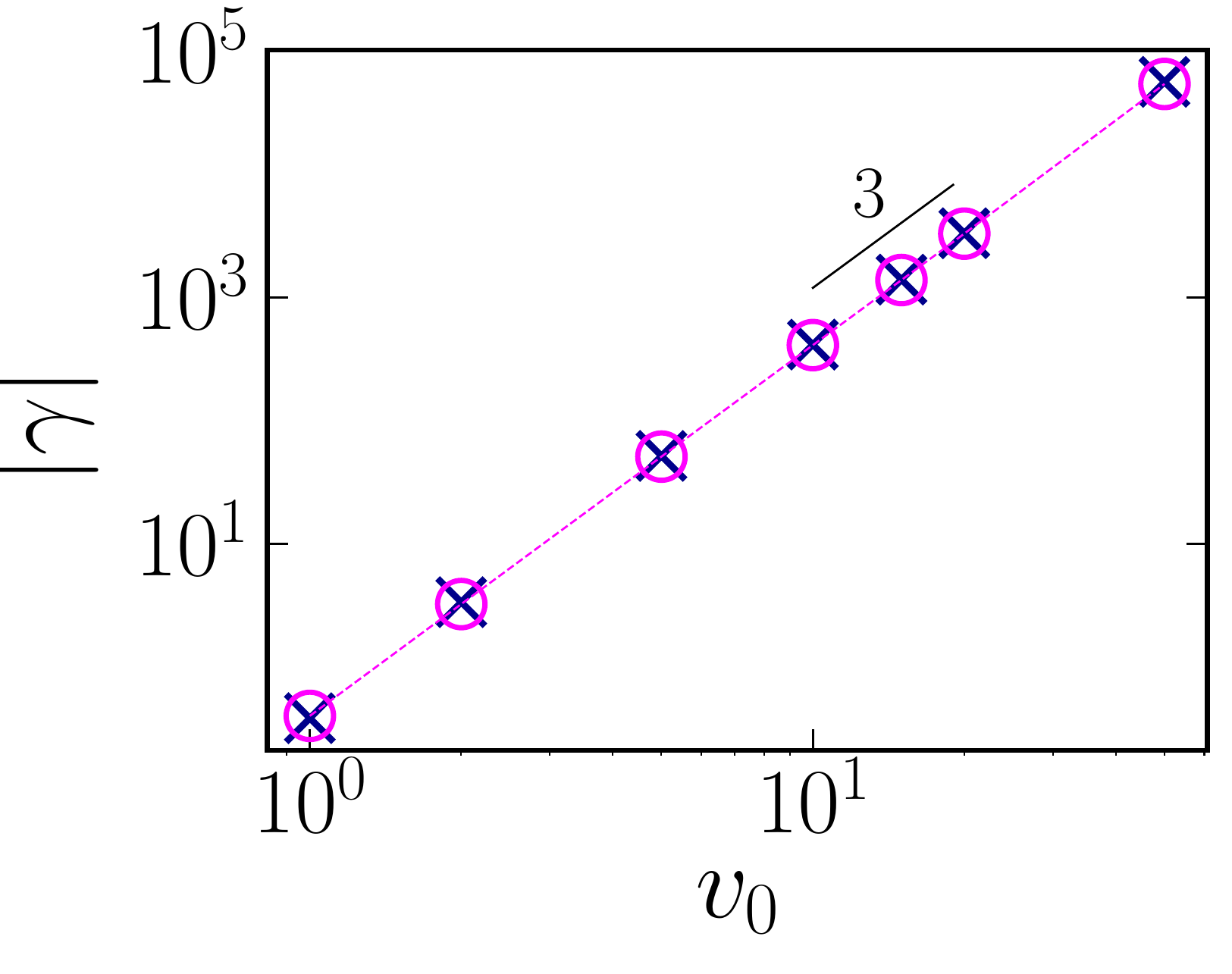}}
  \caption{\textbf{Left:} For hard walls, $\F_i^\mathrm{ext}$ can be decomposed into pressure (red) and
    surface tension (blue) contributions. \textbf{Right:} Surface tension on a hard wall as a function of particle speed $v_0$. The Virial expression~\eqref{eq:KBactive} (magenta circles) coincides with the bulk correlator~\eqref{eq:baregamma} (blue crosses). A fit of the data to Eq.~\eqref{eq:gammascaling} leads to $c\simeq 0.40\pm 0.02$ for ABPs in two dimensions.
    Parameters: $L_x=2L_y=120$,
    $N=1024$, $\mu=1$, $\tau=1$, $\lambda=10^4$.}\label{boundaryforces}
\end{figure}

We consider $N$  active particles in a box of dimensions
$L_x \times L_y$, evolving according to:
\begin{align}
\dot\rr_i=v_0 \uu_i+ \mu \F_i^\mathrm{ext}+\sum_j \mu \F_{j\to i}
\label{eq:dynamics_active_full}
\end{align}
where $\F_i^\mathrm{ext}$ is the force exerted by the walls to confine
 particle $i$. Up to now, we had for clarity considered
non-interacting particles. Our results however extend to particles
interacting via pairwise forces $\F_{j\to i}\equiv -\bn
\Phi(\rr_i-\rr_j)$, a case that we now consider more explicitly. To
develop a Virial approach to surface tension, we introduce a modified
Virial $\mathcal{V}(t)\equiv\frac 1 2
\sum_i\langle\rr_i(t)\cdot\rr_i^*(t)\rangle$, where
$\rr^*_i=(x_i,-y_i)$ are divergence-free vectors. We now use that $\dot {\mathcal{V}}(t)=0$ in the
steady-state to get
\begin{equation}\label{eq:virial1}
  0= v_0 \sum_i \langle \uu_i \cdot \rr_i^* \rangle + \mu \langle \sum_i \F_i^\mathrm{ext}\cdot \rr_i^* \rangle+ \mu \langle \sum_{ij} \F_{j\to i} \cdot \rr_i^* \rangle
\end{equation}
In the limit of hard walls, the external force density along a side of
the box is naturally decomposed into pressure and surface tension
contributions, as depicted in Fig.~\ref{boundaryforces}.  Using this
decomposition, a direct evaluation shows that $\langle \sum_i
\F_i^\mathrm{ext}\cdot \rr_i^* \rangle=\gamma
L_x+\gamma(L_x-L_y)-\gamma L_y+P_b L_x L_y-P_b L_y L_x$.  The pressure
contribution cancels out, since $\rr^*$ is divergenceless, and
Eq.~\eqref{eq:virial1} leads to our modified Virial for the surface
tension:
\begin{equation}\label{eq:virial}
  \gamma = -\frac{1}{2(L_x-L_y)} \sum_i \big\langle \rr_i^* \cdot \Big(\frac{v_0}\mu \uu_i +\sum_{j} \F_{j\to i} \Big) \big \rangle\;.
\end{equation}

Much like the usual Virial formula, the presence of $\rr_i^*$ in
Eq.~\eqref{eq:virial} makes its numerical measurement difficult~\cite{speck2016}. Using that $\partial_t \langle\uu_i\cdot\rr_i^*\rangle=0$ in the steady state, we arrive at
\begin{align}
\tau \langle\uu_i\cdot\rr_i^*\rangle=\mu\langle \F_i^\mathrm{ext}\cdot \uu_i^*\rangle+\sum_j\mu\langle\F_{j\to i}\cdot\uu_i^*\rangle\;,
\label{eq:correlation_angle_rstar}
\end{align}
where we have used $\langle \uu_i\cdot\uu_i^*\rangle=0$ in steady-state,
with $\uu_i^*=(\cos\theta_i,-\sin\theta_i)$. All in all, the surface tension can thus be written in a more numerically-friendly way as
\begin{eqnarray}\label{eq:KBactive}
\gamma=&-\frac{1}{2(L_x-L_y)}\big[v_0 \tau \big(\sum_i\langle\F_i^\mathrm{ext}\cdot\uu_i^*\rangle
+\sum_{ij}\langle\F_{j\to i}\cdot\uu_i^*\rangle\big)\nonumber\\
 &\quad+\frac 1 2 \sum_{ij}\langle\F_{j\to i}\cdot(\rr_i^*-\rr_j^*)\rangle
 \big]
\end{eqnarray}
Equation~\eqref{eq:virial} of course also holds for a passive dynamics
in which it offers an alternative to the Kirkwood-Buff
formula~\cite{kirkwood49}, of which Eq.~\eqref{eq:KBactive} is a
generalization to active systems for a solid-liquid
interface. In the noninteracting limit, dimensional analysis shows the surface tension between an
active gas and a hard wall to be given by
\begin{equation}\label{eq:gammascaling}
  \gamma= -c \frac{v_0^3\tau^2}{\mu}\rho_0\;,
\end{equation}
where $c$ is a dimensionless constant. Figure~\ref{boundaryforces}
shows measurements of the bare surface tension using
Eq.~\eqref{eq:baregamma} and the Virial~\eqref{eq:KBactive} to
coincide, and confirm the scaling of $\gamma$ with $v_0$.

\textit{Conclusion.} We have shown that, as in equilibrium,  surface tension can be defined in active fluids by considering the tangential forces exerted on an interface. At a solid-fluid interface, any anisotropy of the boundary is expected to lead to steady currents that contribute to the force balance. We nevertheless identified a bare, equation-of-state abiding contribution to surface tension, which was shown to share properties with its equilibrium counterpart. We hope this work will trigger new theoretical and experimental endeavours into  interfacial problems in active matter, from the role of currents at phase boundaries to that of wetting and capillarity in multi-phase systems.

\begin{acknowledgments}
  {FvW, JT \& RZ thank K. Mandadapu at UC Berkeley for
hospitality and discussions, JT acknowledges support from ANR
Bactterns, YK acknowledges support from I-CORE Program of the Planning
and Budgeting Committee of the Israel Science Foundation, from an
Israel Science Foundation grant and from the NSF-BSF. JT \& YK
acknowledge support from a joint CNRS-MOST grant. MK acknowledges support from the Alexander von Humboldt Foundation.}
\end{acknowledgments}

\bibliographystyle{apsrev4-1}
\bibliography{biblio_surfaceTension}
%\nocite{*}

\end{document}